arXiv:0812.0518 [cond-mat.mtrl-sci]

# Phonon thermal conduction in graphene: Role of Umklapp and edge roughness scattering

D.L. Nika[1,2], E.P. Pokatilov[1,2], A.S. Askerov[2] and A.A. Balandin[1,3,*]

[1]Nano-Device Laboratory, Department of Electrical Engineering, University of California – Riverside, Riverside, California 92521 USA

[2]Department of Theoretical Physics, Moldova State University, Chisinau, MD-2009, Republic of Moldova

[3]Materials Science and Engineering Program, Bourns College of Engineering, University of California – Riverside, Riverside, California 92521 USA

[*] Corresponding author; electronic address (A.A. Balandin): balandin@ee.ucr.edu ; web: www.ndl.ee.ucr.edu

## Abstract

We investigated theoretically the phonon thermal conductivity of single layer graphene. The phonon dispersion for all polarizations and crystallographic directions in graphene lattice was obtained using the valence-force field method. The three-phonon Umklapp processes were treated exactly using an accurate phonon dispersion and Brillouin zone, and accouting for all phonon relaxation channels allowed by the momentum and energy conservation laws. The uniqueness of graphene was reflected in the two-dimensional phonon density of states and restrictions on the phonon Umklapp scattering phase-space. The phonon scattering on defects and graphene edges has been also included in the model. The calculations were performed for the Gruneisen parameter, which was determined from the *ab initio* theory as a function of the phonon wave vector and polarization branch, and for a range of values from experiments. It was found that the near room-temperature thermal conductivity of single layer graphene, calculated with a realistic Gruneisen parameter, is in the range ~ 2000 – 5000 W/mK depending on the defect concentration and roughness of the edges. Owing to the long phonon mean free path the graphene edges produce strong effect on thermal conductivity even at room temperature. The obtained results are in good agreement with the recent measurements of the thermal conductivity of suspended graphene.

**I. Introduction**

Graphene, a planar single sheet of sp$^2$-bonded carbon atoms arranged in honeycomb lattice, has attracted major attention of the physics and device research communities owing to a number of its unique properties [1-5]. From the practical point of view, some of the most interesting characteristics of graphene are its extraordinary high room temperature (RT) carrier mobility $\mu$, in the range $\mu \sim$ 15000 - 27000 cm$^2$V$^{-1}$s$^{-1}$ [1-2], and recently discovered very high thermal conductivity $K$ exceeding $\sim$ 3080 W/mK [6-7]. The reported values of the thermal conductivity of graphene are on the upper bound of those experimentally found for carbon nanotubes (CNTs) or higher. The outstanding electrical current and heat conduction properties are beneficial for the proposed electronic and thermal management applications of graphene [7]. There exists an inherent ambiguity with the thermal conductivity definition for a single atomic plane due to the uncertainty of the thickness. The latter, together with the fundamanetal science and practical importance of understanding heat conduction in a strictly two-dimensional (2D) system such as graphene, motivated the present theoretical study.

We report a detail theoretical investigation of the thermal conductivity of single-layer graphene (SLG) and compare our results with available experimental data. Using our theoretical formalism we analize the factors, which lead to specific values of the thermal conductivity $K$, and consider their dependence on graphene flake parameters such as width, edge roughness and defect concentration. As it follows from the recent graphene investigation [6-7] and published experimental [8-11] and theoretical [12-16] studies of thermal conduction in CNTs, the heat in graphene should be mostly carried by acoustic phonons rather than by electrons. Experimentally, this conclusion is based on the observation that the contributions of charge carriers to the thermal conductivity, estimated from the Wiedemann-Franz law, is extremely small compared to the overall thermal conductivity [7]. We consider the thermal transport in graphene to be at least partially diffusive. It this sense, it is similar to Klemens' treatment of heat conduction in basal planes of graphite [17] as well as Hone et al. [8] and Kim et al. [9] assumptions in their analysis of thermal conduction in CNTs.

Near RT and above the phonon thermal conductivity is limited by the three-phonon Umklapp processes. The Umklapp-limited thermal conductivity has been studied theoretically in graphite [13, 17-20] and in CNTs [21]. The important observation from these and related works is that the phonon Umklapp processes have negligible effect on the heat flux consisting of the low-energy phonons with the small phonon wave vector $q$. The latter would lead to the extremely high thermal conductivity unless the presence of other phonon scattering mechanisms, e.g. rough boundary scattering, becomes effective in the region of small $q$. For larger $q$ the intensity of Umklapp processes increases and this scattering mechanism starts to dominate in limiting the flux of higher energy phonons. The calculated thermal conductivity depends on the initial (low-$q$) integration region where the main scattering mechanisms are not Umklapp processes. For this reason, even though we are interested in the values of the thermal conductivity near RT and above, we carefully included the boundary and defect scattering into consideration.

Despite the importance of the phonon Umklapp scattering in thermal conduction in semiconductors, the commonly used expressions for the phonon relaxation rates in these processes are approximate [17-22]. Normally, one makes the following simplifications while calculating Umklapp scattering rates: (i) substitution of the phonon velocities with an effective value obtained by averaging over the acoustic phonon polarization branches; (ii) omission of the Umklapp processes characterized by the reciprocal lattice vectors $\vec{b}_i$, which are not parallel to the heat flux direction; (iii) approximate accounting of the phonon selection rules and simplified description of the regions of the allowed phonon transactions in the Brillouin zone (BZ). For conventional semiconductors and nanostructures these simplifications are in many cases justified and lead to results in agreement with experiment at RT [23-26]. The latter is mostly due to the overall strong scattering and weak anisotropy in such systems, which makes the specifics of the Umklapp relaxation channels less important. At the same time, these assumptions become unacceptable for anisotropic materials with very high thermal conductivity where phonon scattering is much weaker. The situation is further complicated for graphitic materials due to a large discrepancy in the reported values of the Gruneisen parameter, which results in substantial differences in the calculated thermal conductivity.

In order to capture the specifics of the phonon heat conduction in graphene, we treated the three-phonon Umklapp scattering rigorously avoiding the common simplifications. Our approach uses the calculated phonon dispersion along all directions in BZ and pertinent reciprocal lattice vectors $\vec{b}_i$. We considered all combinations of the phonon states in the three-phonon Umklapp scattering processes allowed by the momentum and energy conservation laws. The accounting of possible phonon scattering channels was done with help of the *scattering diagrams*, which depict the allowed phonon states in BZ for each relevant phonon scattering process. The rest of the paper is organized as follows. In Section II we describe the calculation of the phonon dispersion. Section III provides details of our method for determining the Umklapp scattering rates of the first (with absorption of the phonon $\vec{q}\,'$) and second (with emission of the phonon $\vec{q}\,'$) kind. In Section IV we calculate the thermal conductivity of graphene and analyze its dependence on the flake width, defect density and temperature. Discussion of the obtained values and comparison with the thermal conductivity in other carbon allotropes are given in Section V. We present our conclusions in Section VI.

## II. Phonon Dispersion in Graphene

In this section we describe the theoretical approach and calculation of the phonon dispersion in SLG. The honeycomb crystal lattice of graphene is presented in Figure 1. The rhombic unit cell, shown as a dashed region, can be defined by two basis vectors $\vec{a}_1 = a(3,\sqrt{3})/2$, and $\vec{a}_2 = a(3,-\sqrt{3})/2$, where $a = 0.142$ nm is the distance between two nearest carbon atoms. The empty and black circles in Figure 1 denote the atoms, which belong to the first and second Bravais lattice, respectively. The atom $1_0$ of the first Bravais lattice is surrounded by three atoms $(\bar{1}_0, \bar{2}, \bar{3})$ of the second Bravais lattice. The inside dashed circle indicate the first interaction sphere, which includes the nearest-neighbor (N) atoms of the atom $1_0$ with the coordinates given by the radius-vectors $\vec{R}(\bar{1}_0;1_0) = a(1,0)$, and $\vec{R}(\bar{2}(\bar{3});1_0) = a(-1,\pm\sqrt{3})/2$. The atoms of the second interaction sphere, shown by a dashed circle with a larger diameter, are denoted as the far-distance-neighbors (F). They belong to the

.

same Bravais lattice as the central atom $1_0$ and defined by the radius-vectors $\vec{R}(1(4);1_0) = \pm a(0,\sqrt{3})$; $\vec{R}(2(5);1_0) = \pm a(-3,\sqrt{3})/2$; and $\vec{R}(3,(6);1_0) = \mp a(3,\sqrt{3})/2$.

In order to find the phonon dispersion in graphene we used the valence force field (VFF) method [27-29]. In this method all interatomic forces are resolved into bond-streching and bond-bending forces. The potential energy for the deformed lattice can be written as [27 – 29]

$$V = V^r + V^{2r} + V^\vartheta + V^{\vartheta\perp} + V^{2\vartheta\perp} + V^{rr}, \qquad (1)$$

where $V^r$ is the stretching potential of the $N$-type interactions, which is given by

$$V^r = \frac{1}{2}\kappa^r \sum_{\bar{i},j}(\delta r_{\bar{i}j})^2. \qquad (2)$$

Here $\delta r_{\bar{i}j}$ is the elongation of the bond between the nearest neighbors $\bar{i}$ and $j$ without a change in the angle between bonds. The stretching potential of the $F$-type interactions (atom interactions within the second sphere shown in Figure 1) is written as

$$V^{2r} = \frac{1}{2}\kappa^{2r} \sum_{i,j}(\delta r_{ij})^2, \quad i,j=1,2,3,4,5,6. \qquad (3)$$

The in-plane bending potential for the $N$-type interactions is given by

$$V^\vartheta = \frac{a^2}{2}\kappa^\vartheta (\sum_{j,\bar{i} \neq \bar{k}}(\delta\vartheta_{\bar{i}j\bar{k}})^2 + \sum_{\bar{j},i \neq k}(\delta\vartheta_{ij\bar{k}})^2) \qquad (4)$$

where $\delta\vartheta_{\bar{i}j\bar{k}}$ is a change of the angle $\vartheta$ between the bonds ($j-\bar{i}, j-\bar{k}$). The out-of-plane bending potential for the $N$-type interactions has the form

$$V^{\vartheta\perp N} = \frac{\kappa^{\vartheta\perp}}{2}\sum_j(\sum_{\bar{i}}u_z(\bar{i}) - 3u_z(j))^2, \qquad (5)$$

where $u_\alpha(i)$ and $u_\beta(\bar{k})$ are the components of the displacement vectors of the atoms in the $i,\bar{k}$ nodes, respectively. The out-of-plane bending potential for the $F$-type interactions is given by

$$V^{2\vartheta\perp} = \frac{\kappa^{2\vartheta\perp}}{2}\sum_j(\sum_i u_z(i) - 3u_z(j))^2, \qquad (6)$$

The higher-order streching-streching interaction is described by the following expression

$$V^{rr} = \kappa^{rr} \sum \delta r_{i\bar{j}} \delta r_{k\bar{j}}, \tag{7}$$

Using the force constants defined as

$$\Phi_{\alpha\beta}(i,j) = \frac{\partial^2 V}{\partial u_\alpha(i) \partial u_\beta(j)}, \tag{8}$$

We can create the following dynamic matrices

$$\begin{aligned}
D^N_{\alpha\beta}(\bar{1}_0 1_0 | \vec{q}) &= \sum_{k=1,2,3} \Phi_{\alpha\beta}(\bar{1}_0, k) e^{i\vec{q}\vec{r}(k)} \\
D^F_{\alpha\beta}(1_0 1_0 | \vec{q}) &= \sum_{k=1,\ldots,6} \Phi_{\alpha\beta}(1_0, k) e^{i\vec{q}\vec{r}(k)} \\
D_{\alpha\beta}(1_0 1_0) &= -D^N_{\alpha\beta}(\bar{1}_0 1_0 | q=0) - D^F_{\alpha\beta}(1_0 1_0 | q=0)
\end{aligned} \tag{9}$$

Taking into account that graphene crystal structure consists of two Bravais lattices, one can define the displacements of the atoms from their equilibrium sites as $u_\alpha(i(\bar{i})) = u_\alpha(1_0(\bar{1}_0)) e^{i\vec{q}\vec{r}(i)}$, where $u_\alpha(1_0)$ and $u_\alpha(\bar{1}_0)$ are the amplitudes of the displacement in the first and second Bravais lattice, correspondingly. Introducing the new variables

$$\begin{aligned}
u(\bar{1}_0) + u(1_0) &\equiv w \\
i(u(\bar{1}_0) - u(1_0)) &\equiv v
\end{aligned}, \tag{10}$$

we obtain a system of six equations

$$\begin{aligned}
\omega^2 w_\alpha &= \sum_\beta (D_{\alpha\beta}(1_0 1_0) + \operatorname{Re} D^N_{\alpha\beta}(1_0, \bar{1}_0 | \vec{q}) + \operatorname{Re} D^F_{\alpha\beta}(1_0, 1_0 | \vec{q})) w_\beta + \\
&+ \sum_\beta (\operatorname{Im} D^N_{\alpha\beta}(1_0, \bar{1}_0 | \vec{q}) - \operatorname{Im} D^F_{\alpha\beta}(1_0, 1_0 | \vec{q})) v_\beta, \quad \alpha, \beta = x, y, z \\
\omega^2 v_\alpha &= \sum_\beta (\operatorname{Im} D^N_{\alpha\beta}(1_0, \bar{1}_0 | \vec{q}) + \operatorname{Im} D^F_{\alpha\beta}(1_0, 1_0 | \vec{q})) w_\beta + \\
&+ \sum_\beta (D_{\alpha\beta}(1_0 1_0) - \operatorname{Re} D^N_{\alpha\beta}(1_0, \bar{1}_0 | \vec{q}) + \operatorname{Re} D^F_{\alpha\beta}(1_0, 1_0 | \vec{q})) v_\beta.
\end{aligned} \tag{11}$$

In these equations, in the limit of small $q$, the vectors $\vec{w}$ describe the acoustic (atoms $1_0$ and $\bar{1}_0$ move in phase) vibrations while the vectors $\vec{v}$ represent the optical (atoms $1_0$ and $\bar{1}_0$ move counter phase) vibrations in graphene. For large $q$, the distinction between the acoustic and optical vibrations, i.e. phonons, becomes approximate.

The eigenfrequencies $\omega_s(\vec{q})$ are found by setting the determinant of the six equations in Eqs. (11) equal to zero. The parameters $\kappa^r, \kappa^{2r}, \kappa^\vartheta, \kappa^{\vartheta\perp}, \kappa^{2\vartheta\perp}, \kappa^{rr}$, which define the interaction potentials (see Eqs. (2-7)) are found from the comparison of the calculated dispersion to the experimental data. The in-plane force constants for graphene and graphite are assumed to be the same although the equations governing lattice vibrations are different allowing one to capture the specifics of 2D system. The required force constants have been determined using the available experimental data for graphite [30-33]. The wave vector $\vec{q}$ is selected within the boundaries of the graphene's first BZ shown in Figure 2. The unit vectors of graphene's reciprocal lattice are given as $\vec{b}_1 = \frac{2\pi}{3a}(1,\sqrt{3})$, $\vec{b}_2 = \frac{2\pi}{3a}(1,-\sqrt{3})$. These vectors, together with their sum, a vector $\vec{b}_3 = \frac{4\pi}{3a}(1,0)$, are indicated in Fig. 2 with $\vec{b}_i = \vec{\Gamma\Gamma}_i$, $i$=1,…,6. The distance between the high-symmetry point $\Gamma$ in the BZ center and the point M in the middle of the honeycomb side is equal to $b_3/2 = 2\pi/(3a)$.

By solving Eqs. (11) we obtained 6 phonon polarization branches, enumerated with an index $s$=1,…, 6, which are shown in Figure 3. These branches are (i) out-of-plane acoustic (ZA) and out-of-plane optical (ZO) phonons with the displacement vector along the Z axis; (ii) transverse acoustic (TA) and transverse optical (TO) phonons, which corresponds to the transverse vibrations within the graphene plane; (iii) longitudinal acoustic (LA) and longitudinal optical (LO) phonons, which corresponds to the longitudinal vibrations within the graphene plane. We simulated the phonon dispersion curves for directions between the high-symmetry $\Gamma$ point and a large number of points on the line $M - K$ (see Figures 2 and 3) with an angle step of $0.1^0$. The latter ensured accuracy in determining the three-phonon selection rules of about ~ 2 %. The graphene dispersion, which we obtained with VFF method, is in excellent agreement with the *ab initio* calculations and other available theoretical and experimental data [33- 35].

## III. Phonon Scattering in Graphene

The Boltzmann equation for the spatially non-uniform phonon distribution can be written as [36-37]

$$(\frac{\partial N}{\partial t})\big|_{drift} + (\frac{\partial N}{\partial t})\big|_{scatt} = 0, \tag{12}$$

where $N$ is the number of phonons in the $(\vec{q}, s)$ mode. Using the standard approach, we can perform differentiation and write the scattering term in the relaxation time $\tau$ approximation, i.e., $(\frac{\partial N}{\partial t})\big|_{Scatt} = -\frac{n}{\tau}$. This leads to the following equation

$$n = -\tau \, (\vec{v} \nabla T) \frac{\partial N_0}{\partial T}, \tag{13}$$

where $n$ is the non-equilibrium part of the phonon distribution function $N = N_0 + n$, $N_0 = 1/(Exp(\hbar\omega/(k_B T)) - 1)$ is the Bose-Einstein distribution function, $\nabla T$ is the temperature gradient and $\vec{v} = \partial \omega / \partial \vec{q}$ is the phonon group velocity. It is well-known that all phonon scattering processes in crystals can be divided into the momentum – destroying processes, which directly constribute to the thermal resistance, and the normal processes, which do not contribute to the thermal resistance but affect the thermal conductivity through re-distribution of the phonon modes [18, 37-39]. Here we follow the Klemens' approach for graphite basal planes [17], and focus on the phonon momentum – destroying scattering processes such as three-thonon Umklapp scattering, point-defect and boundary scattering.

We consider two types of the three-phonon Umklapp scattering processes [37 - 38]. The first type is the scattering when a phonon with the wave vector $\vec{q}$ ($\omega$) absorbs another phonon from the heat flux with the wave vector $\vec{q}'(\omega')$, i.e. the phonon leaves the state $\vec{q}$. Another possibility is when a phonon with the wave vector $\vec{q}''(\omega'')$ decays into two phonons with the wave vectors $\vec{q}(\omega)$ and $\vec{q}'(\omega')$, which corresponds to the phonon coming to the state $\vec{q}$. For this type of scattering processes the momentum and energy conservation laws are written as follows

$$\vec{q} + \vec{q}\,' = \vec{b}_i + \vec{q}\,'' \quad (14)$$
$$\omega + \omega' = \omega''$$

where $\vec{b}_i$, $i = 1, 2, 3$ is one of the vectors of reciprocal lattice (see Figure 2). The processes of the second type are those when the phonons $\vec{q}$ of the heat flux decay into two phonons with the wave vectors $\vec{q}\,'$ and $\vec{q}\,''$ leaving the state $\vec{q}$, or, alternatively, two phonons $\vec{q}\,'(\omega')$ and $\vec{q}\,''(\omega'')$ merge together forming a phonon with the wave vector $\vec{q}(\omega)$, which correspond to the phonon coming to the state $\vec{q}(\omega)$. The conservation laws for the second type of the processes are given as

$$\vec{q} + \vec{b}_i = \vec{q}\,' + \vec{q}\,'', \quad i = 4, 5, 6 \quad (15)$$
$$\omega = \omega' + \omega''.$$

The wave vector $\vec{q}$ of the phonon, which carries heat, is considered to be directed along the $\Gamma \to \Gamma_3$ line (see Figure 2). Thus, for the scattering events of the first type we have $(\vec{q}\vec{b}_i) > 0$, while for the scattering processes of the second type the inequality becomes $(\vec{q}\vec{b}_i) < 0$. The systems of Eqs. (14 – 15) consist of three equations each with 4 unknowns $q'_x, q'_y, q''_x$ and $q''_y$. The phonon dispersions required for solving Eqs. (14 – 15) have been obtained in the previous section using VFF method taking into account the anisotropy of BZ in all directions.

We account for all allowed three-phonon Umklapp processes using the *phonon scattering diagrams*, which represent a set of points, i.e. *curve segments*, $l(q'_x q'_y)$ in BZ, for which the energy and momentum conservation conditions of Eqs. (14 – 15) are met. The representative phonon scattering diagramns are shown in Figure 4 - 6. The captions to these figures indicate the specific phonon relaxation channels. Figure 4 presents the curve segments corresponding to the processes of absorption and decay of a phonon with the wave vector $\vec{q}$ of the LA branch. For example, a segment denoted by index 1 is a set of points in BZ for which the LA phonon decay through the channel LA → ZO + TA is allowed by the concervation laws. Analogously, a segment denoted by index 12 indicates two segments (disconnected sets of points) in BZ for which the LA phonon can convert to TO phonon by absorbing another TA phonon, i.e. LA + TA

→ TO. Overall, the processes with the absorption of all types of phonons, e.g. ZA, TA, LA and ZO, are allowed. The decay of LA phonons can also go via ZA, TA, LA and ZO channels. For the small phonon wave vectors $q<3$ nm$^{-1}$, the most active processes are those that involve absorption of LA and TA phonons. The use of the phonon scattering diagrams for taking the relevant line integrals for the scattering rates is outlined below.

In Figure 5 we show the diagram of the allowed scattering processes for TA phonons. The diagram reads in the same way as the one in Figure 4. These phonons, similar to LA phonons, can interact, i.e. absorb and decay, with ZA, TA, LA and ZO phonons. For the small $q<3$ nm$^{-1}$, the most active processes are those that involve ZA, LA and TA phonons. The upper optical branchs shown in Figure 6 can decay into acoustic, e.g. LA, ZA and TA, and optical, e.g. ZO and TO, phonons. The TO branch can also absorb TA and LA phonons and decay into LA, TA and ZO phonons for the large values of the phonon wave vector $q$. The calculations show that the contribution of all optical phonon branches and ZA modes to the thermal conductivity is negligible due to their small group velocities. At the same time, the out of plane vibrations are important for coupling the phonon modes in the single and, particularly, few-layer graphene and provide scattering phase-space for the three-phonon Umklapp processes. The use of the phonon scattering diagrams allowed us to treat the three-phonon Umklapp processes in graphene accurately without simplifications in order to capture the specifics of the phonon transport in this low-dimensional system.

Let us now consider Umklap processes of the first and second type separately and establish their relative importance in the scattering. For simplicity we assume that the initial heat flux consists of phonons in the given direction $\vec{q} \uparrow\uparrow (\Gamma - M)$ [17 - 18]. Under this assumption we can write $N = N_0 + n$, $N' = N'_0, N'' = N''_0$. For the processes of the first type, the difference in the probability that the phonon comes to or leaves the state with the wave vector $\vec{q}$ is $\sim (N+1)(N'_0+1)N''_0 - NN'_0(N''_0+1) = (N_0+1)(N'_0+1)N''_0 - N_0N'_0(N''_0+1) + n(N''_0 - N'_0) = n(N''_0 - N'_0)$. Taking into account that at equilibrium $n=0$, this equation simplifies to $(N_0+1)(N'_0+1)N''_0 - N_0N'_0(N''_0+1) = 0$. Analogously, for the scattering processes of the second type, the difference in the probability that the phonon comes to or leaves the state with the wave vector $\vec{q}$ through the phonon decay is given as $-(N'_0 + N''_0 + 1)n$.

Using the general expression for a matrix element of the three-phonon interaction [18, 37-38] and taking into account all relevant phonon branches and their dispersion as well as all unit vectors of the reciprocal lattice $\vec{b}_1......\vec{b}_6$, directed from the $\Gamma$ point to the centers of neighboring unit cells (see Figure 2), one obtains for the Umklapp scattering rates

$$\frac{1}{\tau_U^{(I),(II)}(s,\vec{q})} = \frac{\hbar \gamma_s^2(\vec{q})}{3\pi \rho v_s^2(\vec{q})} \sum_{s's'';\vec{b}_i} \iint \omega_s(\vec{q}) \omega'_{s'}(\vec{q}') \omega''_{s''}(\vec{q}'') (N_0[\omega'_{s'}(\vec{q}')] \mp N_0[\omega''_{s''}(\vec{q}'')] + \frac{1}{2} \mp \frac{1}{2}) \times$$
$$\times \delta(\omega_s(\vec{q}) \pm \omega'_{s'}(\vec{q}') - \omega''_{s''}(\vec{q}'')) dq'_l dq'_\perp.$$

(16)

Here $\gamma_s(\vec{q})$ is the mode-dependent Gruneisen parameter, which is determined for each phonon wave vector and polarization branch, $\rho$ is the surface mass density, $q'_l$ and $q'_\perp$ are the components of the vector $\vec{q}'$ parallel or perpendicular to the lines defined by Eqs. (14, 15), correspondingly. The Gruneisen parameter is a tensor of third rank, which determines anharmonicity of the crystal lattice. In general it depends on the phonon state, i.e. branch and wave vector, $(s,\vec{q})$ and temperature. The mode-dependent Gruneisen parameters can be calculated as $\gamma_s(\vec{q}) = -[d\ln(\omega_s(\vec{q}))/d\ln V]$. In this sense, the mode-dependent Gruneisen parameters are a measure of the sensitivity of the phonon frequencies to changes in the system volume. In Eq. (16) the upper signs correspond to the processes of the first type while the lower signs correspond to those of the second type.

Unlike previous studies for conventional bulk and thin-film semiconductors we do not make rough assumption for calculation of the integral in Eq. (16). Instead we use our phonon scattering diamgram technique to account for the selection rules and BZ features. The integrals for $q_l, q_\perp$ are taken along and perpendicular the curve segments, correspondingly, where the conditions of Eqs. (14-15) are met (see Figures 4-6). After integration along $q_\perp$ we obtain the line integral

$$\frac{1}{\tau_U^{(I),(II)}(s,\vec{q})} = \frac{\hbar \gamma_s^2(\vec{q}) \omega_s(\vec{q})}{3\pi \rho v_s^2(\vec{q})} \sum_{s's'';b} \int_l \frac{\pm(\omega_{s''}'' - \omega_s)\omega_{s''}''}{v_\perp(\omega_{s'}')} (N_0' \mp N_0'' + \frac{1}{2} \mp \frac{1}{2}) dq_l'. \qquad (17)$$

The integration in Eq. (17) is carried out along the curve segments $l$ depicted in the phonon scattering diagrams (see Figures 4 – 6).

The combined scattering rate in both types of the three-phonon Umklapp processes for a phonon in the state $(s,\vec{q})$ can be calculated as

$$\frac{1}{\tau_U(s,\vec{q})} = \frac{1}{\tau_U^I(s,\vec{q})} + \frac{1}{\tau_U^{II}(s,\vec{q})}. \qquad (18)$$

One should note here that for the small phonon wave vectors, $q \to 0$, the Umklapp limited phonon life-time $\tau_U \to \infty$. For this reason, the calculation of the thermal conductivity with just Umklapp scattering is not possible without an arbitrary truncation procedure. To avoid unphysical assumptions about the limits of integration we accurately include the phonon scattering on boundaries [37-38]. In the case of graphene, the boundary scattering term correspond to scattering from the rough edges of graphene flakes. No scattering happens from the top and bottom sides of graphene flake since it is only one atomic layer thick and the phonon flux is parallel to the graphene plane. We evaluate the rough edges scattering using the standard equation [37-39]

$$\frac{1}{\tau_B(s,q)} = \frac{v_s(\omega_s)}{d} \frac{1-p}{1+p}. \qquad (19)$$

Here $d$ is the width of the graphene flake, $p$ is the specularity parameter, which depends on the roughness at the graphene edges. The Eq. (19) can be further extended to take into account the dependence of the relaxation time on the direction of the phonon wave vector [39].

Another phonon scattering mechanism, which we take into account is scattering on point defects. The scattering rate for this mechanism can be written as [37]

$$\frac{1}{\tau_{PD}(s,q)} = \frac{S_0 \Gamma}{4} \frac{q_s(\omega_s)}{v_s(\omega_s)} \omega_s^2, \quad (20)$$

where $S_0$ is the cross-section area per one atom of the lattice and $\Gamma$ is the measure of the strength of the point defect scattering. In the case when the point-defect scattering is only due to the difference in the mass of atoms it is given as $\Gamma = \sum_i f_i (1 - M_i / \bar{M})^2$, where $f_i$ is the fractional concentration of impurity atoms, $M_i$ is the mass of $i$th impurity atom of defect and $\bar{M} = \sum_i M_i f_i$ is the average atomic mass. This scattering term is non-zero in natural carbon materials even if there are no defects or impurities owing to the presence of the isotopic scattering. After the scattering rates in separate relaxation processes are written one can express the combined phonon relaxation rate as

$$\frac{1}{\tau_{tot}(s,q)} = \frac{1}{\tau_U(s,q)} + \frac{1}{\tau_B(s,q)} + \frac{1}{\tau_{PD}(s,q)}. \quad (21)$$

### IV. Thermal Conductivity of Graphene

The heat flux along a graphene flake can be calculated according to the expression [37]

$$\vec{W} = \sum_{s,\vec{q}} \vec{v}(s,\vec{q}) \hbar \omega_s(\vec{q}) N(\vec{q}, \omega_s(\vec{q})) = \sum_{s,\vec{q}} \vec{v}(s,\vec{q}) \hbar \omega_s(\vec{q}) n(\vec{q}, \omega_s), \quad (22)$$

where $\vec{v}\hbar\omega$ is the energy carried by one phonon and $N(\omega, \vec{q})$ is the number of phonons in the flux. Substituting Eq. (13) into Eq. (22) one has

$$\vec{W} = -\sum_{\beta} (\nabla T)_{\beta} \sum_{s,\vec{q}} \tau_{tot}(s,\vec{q}) v_{\beta}(s,\vec{q}) \frac{\partial N_0(\omega_s)}{\partial T} \vec{v}(s,\vec{q}) \hbar \omega_s(\vec{q}) \qquad (23)$$

Using the macroscopic definition of the thermal conductivity

$$W_{\alpha} = -\kappa_{\alpha\beta} (\nabla T)_{\beta} h L_x L_y, \qquad (24)$$

we obtain the following expression for the thermal conductivity tensor

$$\kappa_{\alpha\beta} = \frac{1}{hL_xL_y} \sum_{s,\vec{q}} \tau_{tot}(s,\vec{q}) v_{\alpha}(s,\vec{q}) v_{\beta}(s,\vec{q}) \frac{\partial N_0(\omega_s)}{\partial T} \hbar \omega_s(\vec{q}). \qquad (25)$$

Here $L_x = d$ is the sample width (graphene flake width) and $L_y$ is the sample length. We can write the diagonal element of the thermal conductivity tensor, which corresponds to the phonon flux along the temperature gradient, as

$$\kappa_{\alpha\alpha} = \frac{1}{hL_xL_y} \sum_{s,\vec{q}} \tau_{tot}(s,\vec{q}) v^2(s,\vec{q}) \cos^2 \varphi \frac{\partial N_0(\omega_s)}{\partial T} \hbar \omega_s(\vec{q}). \qquad (26)$$

Finally, making a transition from the summation to integration, we take into account the two-dimensional density of phonon states and obtain the expression for the scalar thermal conductivity

$$\kappa = \frac{1}{4\pi kT^2 h} \times$$
$$\times \sum_{s=1...6} \int_0^{q_{max}} \{ [\hbar \omega_s(q) v_s(q)]^2 \tau_{tot}(s,q) \frac{Exp(\hbar \omega_s(q)/kT)}{(Exp(\hbar \omega_s(q)/kT) - 1)^2} q \} dq \qquad (27)$$

Using Eq. (27) we calculated the thermal conductivity of graphene as a function of temperature. The important features of our theoretical approach are that (i) the actual phonon dispersion of graphene, obtained with VFF method was used; (ii) all phonon polarization branches have been taken into account in determining the thermal conductivity; (iii) and three-phonon Umklapp scattering processes were considered without simplifications. The 2D-nature of the material system was reflected via the proper phonon density of states (DOS) arising from the actual phonon dispersion and three-phonon Umklapp scattering space determined with *scattering diagrams*. The phonon DOS and relation rates enter the expressions for the thermal conductivity.

First, we examined the sensitivity of the thermal conductivity to the value of the Gruneisen parameter (see Figure 7). The phonon mode-dependent Gruneisen parameter for graphene as functions of the phonon wave vector and phonon polarization, $\gamma_s(q)$, was taken from the ab initio calculations of Mounet and Marzari [34] who determined it for all six phonon polarization branches [29]. We used the mode-dependent $\gamma_s(q)$ in most of our calculations so that the phonon interactions (decay and absorption) were accounted with the Gruneisen parameter specific for an each given phonon state $(s, \vec{q})$. In the theory of conventional semiconductors it is more common to treat Gruneisen parameter as a scalar constant, $\gamma$, characteristic for a given material and independent of the phonon mode or temperature [17-19]. There is substantial discrepancy in the reported values of the Gruneisen parameter for graphene, graphite and CNTs [17, 31, 40 – 44]. The Gruneisen parameters as high as $\gamma = 2.0$ have been suggested in Ref. [17] and measured to be as low as 1.06 [31, 43] and 1.11 [45] for graphite basal planes, which are considered to be similar to graphene. In Ref. [34] the lowest characteristic value for Gruneisen parameter was determined to be 0.8. Moreover, it was suggested that the in-plane Gruneisen parameter for graphite basal planes decreases substantially with increasing temperature [46-47]. The latter suggests that the sample heating during the measurements may lead to a reduction in the Gruneisen parameter of graphene and increase of graphene's thermal conductivity [48]. For this reason, the thermal conductivity curves shown in Figure 7 were calculated for a range of realistic Gruneisen parameters. Some of them came from the first-principle calculations, others – from the measurements for graphene or closely related material systems such as CNTs or graphite.

One can see from Figure 7, that the calculated values of the thermal conductivity at RT may vary from ~3000 W/mK ($\gamma=2$) to ~6500 W/mK ($\gamma=0.8$). The results for the mode-dependent Gruneisen parameter $\gamma_s(q)$ are in the middle giving the RT thermal conductivity of ~4000

W/mK. The choice of $\gamma$ produces a rather pronounced effect but not as strong as one would expect for the pure Umklapp-limited phonon thermal conductivity. The phonon life-time in three-phonon Umklapp process $\tau_U$ is proportional to $1/\gamma^2$ (see Eq. (17)). Our results indicate that the variation in $\gamma$ value in the range $0.8 \le \gamma \le 2$ leads to an average change of the thermal conductivity in the considered temperature interval by a factor of ~2. The dependence of the thermal conductivity on $\gamma$ is weakened due to the presence of other scattering mechanisms. For comparison, we also show the experimental data point with the vertical error bar indicating the data spread for a set of graphene flakes [6-7]. Although the measurements, reported in Refs. [6-7], were conducted at RT, there was a substantial heating induced at the center of graphene flake so that the measured thermal conductivity can be referred to the temperature of about 350K [48]. It follows from Figure 7 that the thermal conductivity calculated with the mode-dependent Gruneisen paremeter $\gamma_s(q)$ from Ref. [34] gives the best agreement with the measurements [6-7].

The calculated high values of the thermal conductivity suggest that the phonon mean free (MFP) in graphene is long even at RT. The latter may result in the strong dependence of the thermal conductivity on the width $d$ of the flake and roughness of its edges since the phonon boundary scattering starts to play a prominent role when $d$ is comparable to MFP. Figure 8 shows the thermal conductivity as a function of temperature as the width of the flake varies from $d$=3 to 9 μm. In order to analyze the effect of the graphene flake edges one has to define the speculariry parameter $p$ (see Eq. (19)). For the ideally specular phonon reflection from the edges ($p$=1) the boundary scattering does not add to the thermal resistance. For Figure 8 the specularity parameter was taken to be $p$=0.9. This value was determined from the geometry considerations following Ziman's method [39] and available scanning electron microscopy data for graphene thermal experiments [6-7, 48]. The variation of the flake width from $d$=3 to 9 μm leads to the RT thermal conductivity change by about a factor of 1.8. Several flakes examined in Refs. [6-7] had the width of about 5 μm. It is interesting to note that the experimental data point indicated in Figure 8 is closest to the theoretical curve obtained for $d$=5 μm. The sensitivity of the calculated RT thermal conductivity of graphene to the width of the flakes and roughness of their edges may explain a rather wide range of the experimentally measured values.

The strong effect of the specularity parameter and the flake width on the thermal conductivity near RT can be attributed to the very large phonon mean free path (MFP) and the fact that large fraction of the heat is carried by the low-energy phonons, which are strongly affected by the boundary scattering. The phonon MFP in graphene at RT extracted from the experimental data was about 775 nm [7]. Since it is rather close to the width of the examined graphene flakes one can expect on effect on thermal conductivity even at RT. In this regard, our calculations are in agreement with the experimental findings. Another conclusion from these results is that since edges produce a very strong effect on the thermal conductivity of graphene one needs to either determine the speculariry parameter very accurately from the experiment or develop an atomistic model for the graphene flake edge scattering.

Figure 9 presents the calculated thermal conductivity of graphene over a wide temperature range. In the low-temperature limit the thermal conductivity increases rapidly with increasing temperature as the number of phonons increase. The decrease in the thermal conductivity with temperature, which starts around 80 K is due to the growing strength of the Umklapp scattering processes. It is interesting to note that the thermal conductivity in the low-temperature limit is proportional to $T^2$. Indeed, for the curve calculated with $p=0.9$, the ratio $K(T=80\ K)/K(T=50\ K) = 2.50$, while the ratio of the temperatures squared is $(80/50)^2=2.56$. This is a manifestation of the 2D nature of graphene. In bulk the low-temperature thermal conductivity is proportional to $\sim T^3$. The low-temperature $T^2$-dependence of the graphene thermal conductivity can be obtained from Eq. (27) analytically through a standard procedure by extending the upper integration limit to infinity. The difference in the temperature dependence between the 2D graphene and bulk materials is related to the different phonon density of states. Small deviation from $T^2$ dependence in our case is explained by the fact that the considered temperature is not low enough (~50-80 K) and by the presence of other scattering mechanisms, e.g. phonon – boundary scattering. The latter is confirmed by the growing deviation as $p$ decreases farther down from unity. For example, the deviation from $T^2$ dependence is somewhat larger for the curve with $p=0.8$ shown in Figure 7.

One should note here that our calculations included the isotopic effect through the point-defect scattering term (see Eq. (20)) on the thermal conductivity. The two stable isotopes of carbon, $^{12}$C and $^{13}$C, have natural abundances of 98.9% and 1.1%, respectively. The overall trend obtained in our calculations is in line with the results reported for another allotrope of carbon, diamond [49-50], although the dependence is weaker. A more accurate accouting of normal phonon processes might be needed to determine the exact scale of the isotop effects in graphene. From our calculations we were also able to estimate the contributions of different phonon polarizations (see Figure 3) to thermal transport. At temperature T=100 K, TA modes transfer about ~28.5% of heat while LA modes carry about 71.0%. The remaining ~0.5% are the contributions of all other phonon polarizations (vibrational modes) of graphene. As the temperature increases above RT, the relative contributions of different phonons change. At T=400 K, TA and LA modes carry ~49% and 50% of heat, correspondingly. The rest of the modes, including out-of-plane phonons, carry ~1% of heat in graphene.

**V. Comparison with Other Carbon Allotropes**

In this section we rationalize the obtained values of the thermal conductivity and compare them with those for other carbon allotropes. The first thing to note is that there is inherent ambiguity in the definition of the thermal conductivity of an individual atomic plane due to the uncertainty of the thickness. In our calculation we used $h$=0.35 nm as a thickness of an individual graphene layer. This value corresponds to the interlayer spacing of graphite, i.e. bond length, and is usually taken as a thickness of CNT as well. The only reported experimental studies of the thermal conductivity of graphene [6-7] used the same value, which facilitated our comparison. At the same time, this definition is not unique [51]. If one introduces the thickness directly from the interatomic potential it can be as low as $h$=0.06 nm [51]. The alternative approach, which uses proper values of the Young's modulus and tensile strength as the starting points, leads to another thickness of graphene giving the values as large as $h$=0.69 nm [52]. It is clear that depending on the chosen definition one may have substantial discrepancy in the calculated or measured thermal conductivity. The important conclusion, which follows from this ambiguity, is that one can rather accurately compare the thermal conductivity of graphene with that of CNT provided the same thickness definition was used. A comparison with bulk allotropes like diamond or amorphous carbons is less accurate and has to be treated with reservations. For this reason we mostly focus on comparison with CNTs. The values of the thermal conductivity of few-layer graphene, involving at least several atomic planes, are much less ambiguous.

In Table I we summarize experimental RT thermal conductivity of graphene (the only measurement reported to date) and CNTs. One can see that there is substantial discrepancy in the reported data from as low as 1500 W/mK [53] to as high as 7000 W/mK [10]. The highest thermal conductivities obtained in the experiments were attributed to completely *ballistic* transport regime achieved in some CNTs. The mostly quoted values for CNTs, which we consider to be commonly accepted, are 3000 – 3500 W/mK [9, 11]. In this sense, the measured thermal conductivity of graphene [6-7] is on the upper bound of what was reported for CNTs. Our modeling results are in good agreement with the experimental data of Refs [6-7] although there is rather substantial range of the obtained values due to the ambiguity in the Gruneisen parameter and strong dependence on the graphene flake width. In our calculations, the best match to the measured thermal conductivity [6-7] is obtained with the mode-dependent Gruneisen parameter from the first-principle theory of Ref. [34]. The "recommended value" of the in-plane (basal plane) thermal conductivity of high quality pyrolitic graphite compiled on the basis of many experimental reports is 2000 W/mK at RT [54]. Many commercial samples have the in-plane thermal conductivity in the range from ~500 to 1700 W/mK. At the same time, the RT experimental values as high as 3000 W/mK were reported for basal planes of graphite [55]. There is also a clear continuing trend that the thermal conductivity of graphite increases with the improvements in processing technology as the crystallinity improves, e.g. the granular size becomes larger [56]. The use of the suspended graphene samples in the experiments helped to select the best crystalline graphene layers, with the highest thermal conductivity, since the lower quality samples collapse.

The theoretical treatment of the thermal conductivity of the graphite basal planes, graphene and CNTs can be roughly divided into two groups. The first is the molecular dynamics (MD) simulations, which usually utilize the Tersoff – Brenner potential for C-C interactions and Green – Kubo relation for the extraction of the thermal conductivity from the heat current correlation functions [12-16]. The second is based on the solution of the Boltzmann's equations with the phonon scattering rates determined from the perturbation theory or fitted to the experimental data. The second approach is commonly referred to as Callaway – Klemens [17-19, 37-38, 57]. Its accuracy can vary depending on the assumptions, particularly, in the treatment of Umklapp processes. Table II gives theoretical RT thermal conductivity for CNTs obtained with MD calculations [12-14] as well as a data point for the basal graphite plane obtained using the Callaway – Klemens model [17]. The cited calculations for CNTs also had results or extrapolations for the thermal conductivity in graphene. Mostly the conclusion, based on MD calculations, was that the thermal conductivity in graphene should be somewhat higher than that in CNTs near RT [13-14]. Overall, the theoretical data scatter for CNTs is large, giving values from 1500 W/mK to 6600 W/mK at RT.

Berber et al. [13] used the equilibrium and non-equilibrium MD simulations to obtain some of the highest thermal conductivities of graphene with RT values as high as 6600 W/mK. Based on their MD simulations and Green – Kubo expression for the thermal conductivity, the authors concluded that once graphene layers are stacked in bulk graphite, the interlayer interactions quench the thermal conductivity by "nearly one order of magnitude" [13]. This fact was checked by simulating the thermal conductivity of the high-quality graphite along basal planes. The quenching may correspond to the increased Umklapp scattering in bulk graphite as the scattering phase-space becomes larger. Thus, we expect that addition of the atomic layers will lead to graduale reduction of the thermal conductivity of few-layer graphene and convergence of the thermal conductivity with the in-plane bulk graphite value.

The sound velocity, or phonon group velocity, used in the calculations is a parameter, which may strongly affect the final result. For example, Klemens obtained rather high thermal conductivity of graphite basal planes on the order of $\kappa \sim 4400$ W/mK with the very large average Gruneisen parameter $\gamma = 2.0$ [17]. The obtained high thermal conductivity was likely a result of the overestimated sound velocities used in his calculation: $v(LA) = 23.6$ km/s, $v(TA) =$ 15.9 km/s. In our calculations we used the phonon velocities from the dispersion obtained with the VFF method. These velocities closely coincided with the recent experimental data obtained for in-plane graphite velocities [33]. The measured longitudinal and transverse velocities were $v(LA)$ = 21.3 km/s and $v(TA)$ = 13.6 km/s. Since the Umklapp-limited thermal conductivity is proportional to the phonon velocity to the power of four (see Eqs. (16) and (17) in conjunction with Eq. (27)) the above mentioned difference in the velocities leads to the thermal conductivity change by a factor of ~1.7.

It is interesting to note that carbon allotropes reveal an extremely wide range of the thermal conductivities: from the lowest values in the "thermal insulators" such as amorphous carbon (*a*-C) to the highest in the "heat superconductors" such as graphene and CNTs. The peak of the thermal conductivity of the best bulk crystal heat conductor – diamond – reaches the value of ~ 41000 W/mK at T=104 K (obtained for 99.9% pure sample) [58]. The RT thermal conductivity of a typical Type IIa diamond is in the range 2000 – 2500 W/mK [59]. The thermal conductivity of CNTs and graphene are even higher (see Table I and II). From the other side, *a*-C, diamond-like carbon (DLC) or nanocrystalline diamond (*n*-D) are very bad conductor of heat with the RT thermal conductivities in the range 0.1 – 10 W/mK [60-62]. According to the measurements and theory the low thermal conductivity of disordered carbon allotropes is associated with large disorder or phonon scattering on the polycrystalline granular interfaces [62]. Another observation is the strong effect produced by graphene edges on the phonon transport. The latter is yet another manifestation of the importance of graphene edges in addition to the edge effect on the electric current and spin transport in graphene [63].

## VI. Conclusions

We studied theoretically the phonon thermal conductivity of single layer graphene. The model and calculation procedure did not use any *hidden* fitting parameters. The only parameters fitted to experimental data were force constants in the VFF method used to determin the phonon dispersion for all polarizations and crystallographic directions in graphene. The obtained dispersion is in good agreement with the calcutaions by other methods and available experimental data. The three-phonon Umklapp processes were treated exactly, without common simplifications, using an accurate phonon dispersion and accouting for all phonon relaxation channels allowed by the momentum and energy conservation laws. The latter was made possible with the help of the phonon *scattering diagram* technique. The unique 2D nature of graphene was reflected in the proper phonon density of states and restrictions on the phonon Umklapp scattering space. Most of the calculations were performed for a mode-dependent Gruneisen parameter taken from the *ab initio* theory. We also examine the variations in the thermal conductivity calculated with differerent values of the Gruneisen parameter taken from experiments. The results of the calculations have been compared with the available experimental and theoretical results for graphene and CNTs. Possible reasons for high thermal conductivity were discussed in details.

*Note Added in Proof*

We became aware of the direct measurement of the Gruneisen parameter of graphene for the optical ($E_{2g}$) phonons [64]. The measured values were in excellent agreement with the calculated mode-dependent Gruneisen parameter [34]. This suggests that the Gruneisen paremeters [34] for acoustic phonons, which we used in our calculations, are also accurate.

We recently learned about two theoretical studies. Jiang et al [65] calculated thermal conductance of graphene in the ballistic limit. The authors obtained a very high value, which translates to the thermal conductivity in excess of 6600 W/mK. It is expected for the ballistic transport regime and, thus, in line with our calculations. Kong et al [66] obtained the thermal conductivity of 2200 W/mK. We note however that the calculation in Ref [66] utilizes an unspecified truncation procedure for the low-frequency phonons, which is completely unwarranted for graphene, and essentially determines the final "bulk" graphite result (see our Comment [67] on this paper for details).

*Acknowledgements*

The work at UCR was supported, in part, by DARPA – SRC Focus Center Research Program (FCRP) through its Center on Functional Engineered Nano Architectonics (FENA) and Interconnect Focus Center (IFC), and by AFOSR award A9550-08-1-0100 on the Electron and D.L. Nika, E.P. Pokatilov, A.S. Askerov and A.A. Balandin, Phonon thermal conduction in graphene: Role of Umklapp and edge roughness scattering, *Phys. Rev. B* (2009) – Editors' Suggestion
.

Phonon Engineered Nano and Heterostructures. A.A.B. acknowledges useful discussions on phonons in graphene with Drs. K.S. Novoselov (University of Manchester) and N. Bonini (MIT). The authors are grateful to Drs. M. Makeev (NASA Ames) and A. Khitun (UCLA) for critical reading of the original version of the manuscript.

**Figure Captions**

Figure 1: Graphene crystal lattice. The rombic unit cell is shown as a shaded region.

Figure 2: Reciprocal lattice of graphene.

Figure 3: Phonon dispersion in graphene calculated using the VFF method.

Figure 4: Three-phonon scattering diagrams used for accounting of the LA phonon Umklapp scattering processes with participation of the following phonons: (1) LA → ZO + TA, $q = 11.8$ nm$^{-1}$ ($\Gamma_4$); (2) LA → ZO + ZA, $q = 11.8$ nm$^{-1}$ ($\Gamma_4$); (3) LA → TA + ZO, $q = 11.8$ nm$^{-1}$ ($\Gamma_4$); (4) LA → ZO + ZA, $q = 11.8$ nm$^{-1}$ ($\Gamma_5, \Gamma_6$); (5) LA + ZA → LO, $q = 13.25$ nm$^{-1}$ ($\Gamma_3$); (6) LA + ZA → TO, $q = 11.8$ nm$^{-1}$ ($\Gamma_3$); (7) LA + ZA → LA, $q = 8.8$ nm$^{-1}$ ($\Gamma_3$); (8) LA + ZO → TO, $q = 5.8$ nm$^{-1}$ ($\Gamma_3$); (9) LA + LA → TO, $q=1.2$ nm$^{-1}$ ($\Gamma_3$); (10) LA + TA → LA, $q = 5.8$ nm$^{-1}$ ($\Gamma_3$); (11) LA + ZA → TA, $q = 4.3$ nm$^{-1}$ ($\Gamma_1, \Gamma_2$); (12) LA + TA → TO, $q = 1.2$ nm$^{-1}$ ($\Gamma_1, \Gamma_2$).

Figure 5: Three-phonon scattering diagrams used for accounting of the TA phonon Umklapp scattering processes with participation of the following phonons: (1) TA → ZA+ZA, $q = 10.27$ nm$^{-1}$ ($\Gamma_4$); (2) TA → ZA + ZA, $q = 11.8$ nm$^{-1}$ ($\Gamma_4$); (3) TA → ZA + ZA, $q = 13.25$ nm$^{-1}$ ($\Gamma_4$); (4) TA → TA + ZA, $q = 11.8$ nm$^{-1}$ ($\Gamma_5, \Gamma_6$); (5) TA → ZA + ZO, $q = 13.25$ nm$^{-1}$ ($\Gamma_4$); (6) TA +

TA → LO, $q = 11.8$ nm$^{-1}$ ($\Gamma_3$); (7) TA + ZA → TA, $q = 8.8$ nm$^{-1}$ ($\Gamma_3$); (8) TA + TA → LA, $q = 8.8$ nm$^{-1}$ ($\Gamma_3$); (9) TA + TA → LO, $q=11.8$ nm$^{-1}$ ($\Gamma_3$); (10) TA + ZA → ZA, $q = 4.3$ nm$^{-1}$ ($\Gamma_1,\Gamma_2$); (11) TA + LA → LO, $q = 2.8$ nm$^{-1}$ ($\Gamma_3$).

Figure 6: Three-phonon scattering diagrams used for accounting of the TO and LO phonon Umklapp scattering processes with participation of the following phonons: (1) TO → TA+ZO, $q = 5.8$ nm$^{-1}$ ($\Gamma_4$); (2) LO → ZA + ZA, $q = 5.8$ nm$^{-1}$ ($\Gamma_4$); (3) TO → TA + ZO, $q = 5.8$ nm$^{-1}$ ($\Gamma_4$); (4) TO → TA+TA, $q = 11.8$ nm$^{-1}$ ($\Gamma_4$); (5) LO → ZO + TA, $q = 5.8$ nm$^{-1}$ ($\Gamma_5,\Gamma_6$); (6) TO → TA + ZO, $q = 11.8$ nm$^{-1}$ ($\Gamma_4$); (7) TO → LA+TA, $q = 11.8$ nm$^{-1}$ ($\Gamma_4$); (8) LO → TA + LA, $q = 13.2$ nm$^{-1}$ ($\Gamma_4$); (9) LO → TA + ZO, $q = 11.8$ nm$^{-1}$ ($\Gamma_5,\Gamma_6$); (10) TO → TA + ZO, $q = 5.8$ nm$^{-1}$ ($\Gamma_5,\Gamma_6$).

Figure 7: Thermal conductivity of graphene as a function of temperature plotted for different values of Gruneisen parameter. The results were calculated for the graphene flake with the width of 5 μm and specularity parameter p=0.9. An experimental data point after Refs. [6-7] is also shown for comparison.

Figure 8: Thermal conductivity of graphene as a function of temperature shown for different graphene flake widths. The results were calculated for the specularity parameter p=0.9. An experimental data point after Refs. [6-7] is also shown for comparison. The best agreement is observed for the curve calculated with the mode-dependent Gruneisen parameter [34].

Figure 9: Thermal conductivity of graphene over a wide temperature range calculated for the graphene flake with the width of 5 μm and mode-dependent Grunesien parameter [34]. The results are obtained for two values of the specularity parameter p=0.9 and point-defect scattering strength Γ. An experimental data point after Refs. [6-7] is also shown for comparison.

**Table I: Experimental Thermal Conductivity of Graphene and CNTs near Room Temperature**

| Sample | K (W/mK) | Method | Comments | Reference |
|---|---|---|---|---|
| graphene | ~3080 – 5300 | optical; non-contact | single atomic layer; diffusive | Balandin et al.[a] |
| MW-CNT | >3000 | electrical; self-heating | individual; diffusive transport | Kim et al.[b] |
| SW-CNT | ~3500 | electrical; self-heating | individual; boundary-limited | Pop et al.[c] |
| SW-CNT | 1750 – 5800 | thermocouples | bundles; diffusive transport | Hone et al.[d] |
| SW-CNT | 3000 – 7000 | thermocouples | individual; ballistic transport | Yu et al.[e] |
| CNTs | 1500 – 2900 | electrical; nano-sensors | individual | Fujii et al.[f] |
| bulk graphite | 500 – 2000; max > 2000 | variety | in-plane (basal); high quality | Ho et al.[g] |

[a]Reference 6  
[b]Reference 9  
[c]Reference 11  
[d]Reference 8  
[e]Reference 10  
[f]Reference 53  
[g]Reference 55

**Table II: Theoretical Thermal Conductivity of Graphene and CNTs near Room Temperature**

| Sample | K(W/mK) | Method | Comments | Reference |
|---|---|---|---|---|
| graphene | 2000 - 5000 | Klemens – type with accurate dispersion | diffusive; exact Umklapp; strong edge dependence | Nika et al.[a] |
| CNT | ~6600 | MD and Green-Kubo for (10,10) nanotube | predicted higher K for graphene than for CNTs | Berber et al.[b] |
| CNT | ~2980 | MD and Green-Kubo | strong defect dependence | Che et al.[c] |
| CNT | 1500 - 2500 | MD with Tersoff-Brener | comparable with graphene | Osman et al.[d] |
| bulk graphite | ~2000 | Callaway – Klemens type; no dispersion | basal plane; approximate Umklapp | Klemens et al.[e] |

[a]Nika, Pokatilov, Askerov and Balandin (this work)
[b]Reference 13
[c]Reference 12
[d]Reference 14
[e]Reference 17

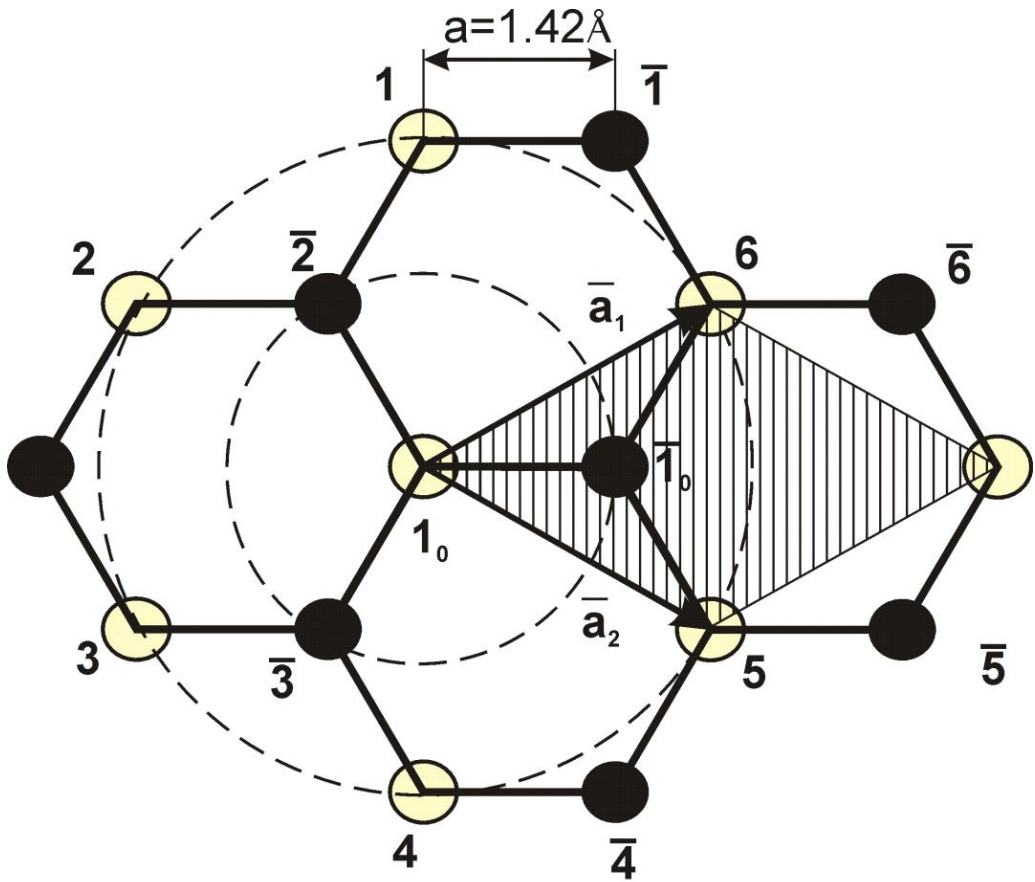

Fig. 1 of 9. Nika et.al.

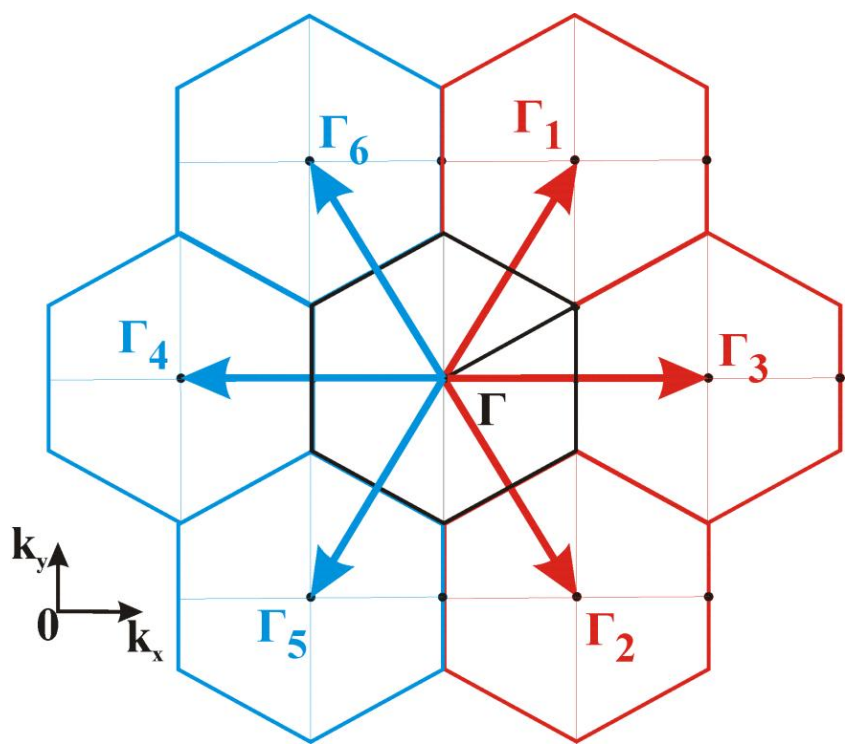

Fig. 2 of 9. Nika et.al.

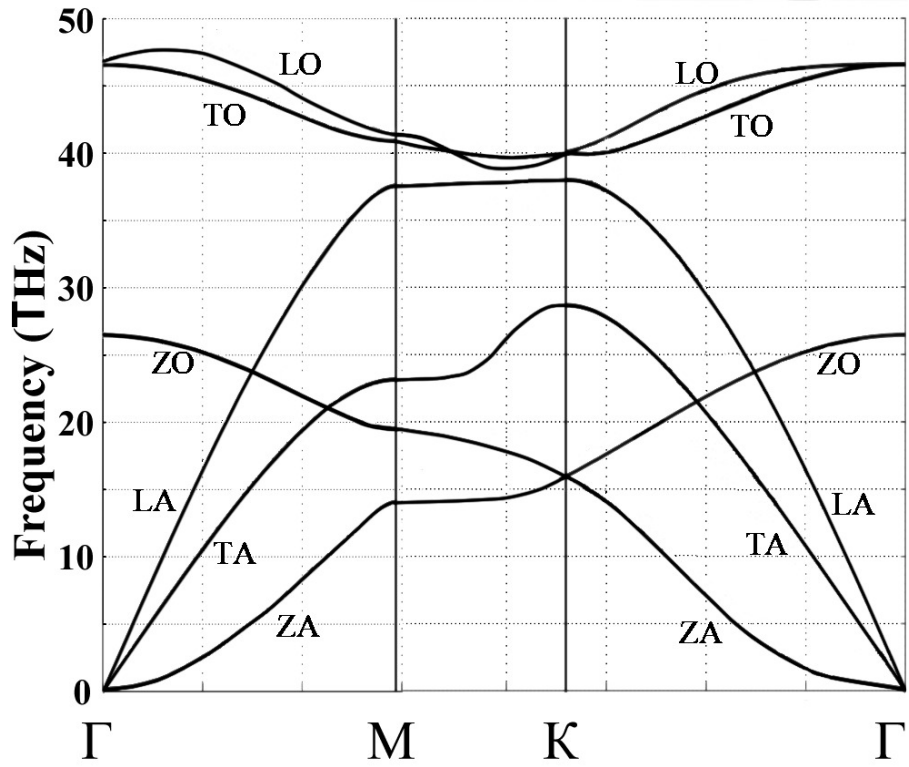

Fig. 3 of 9. Nika et.al.

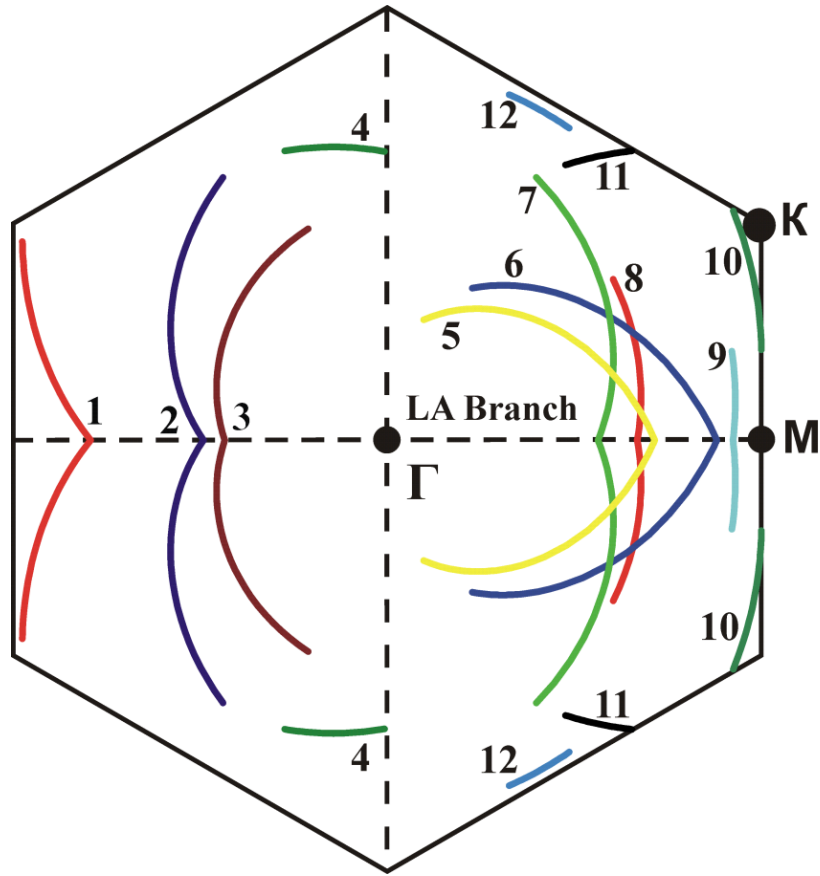

Fig. 4 of 9. Nika et.al.

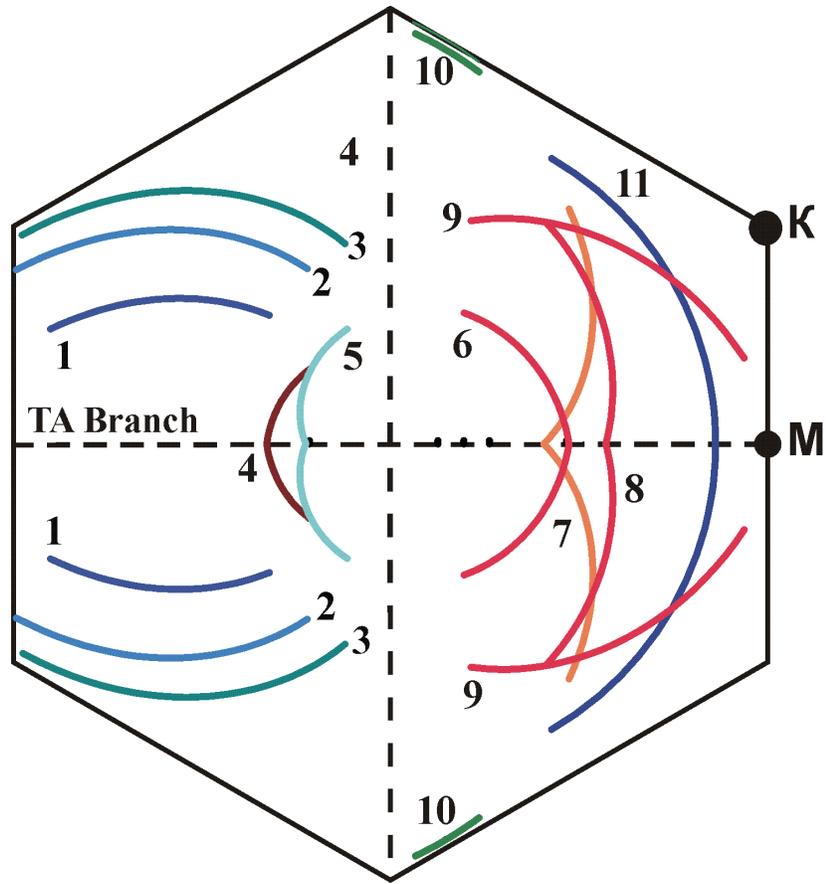

Fig. 5 of 9. Nika et.al.

Fig. 6 of 9. Nika et.al.

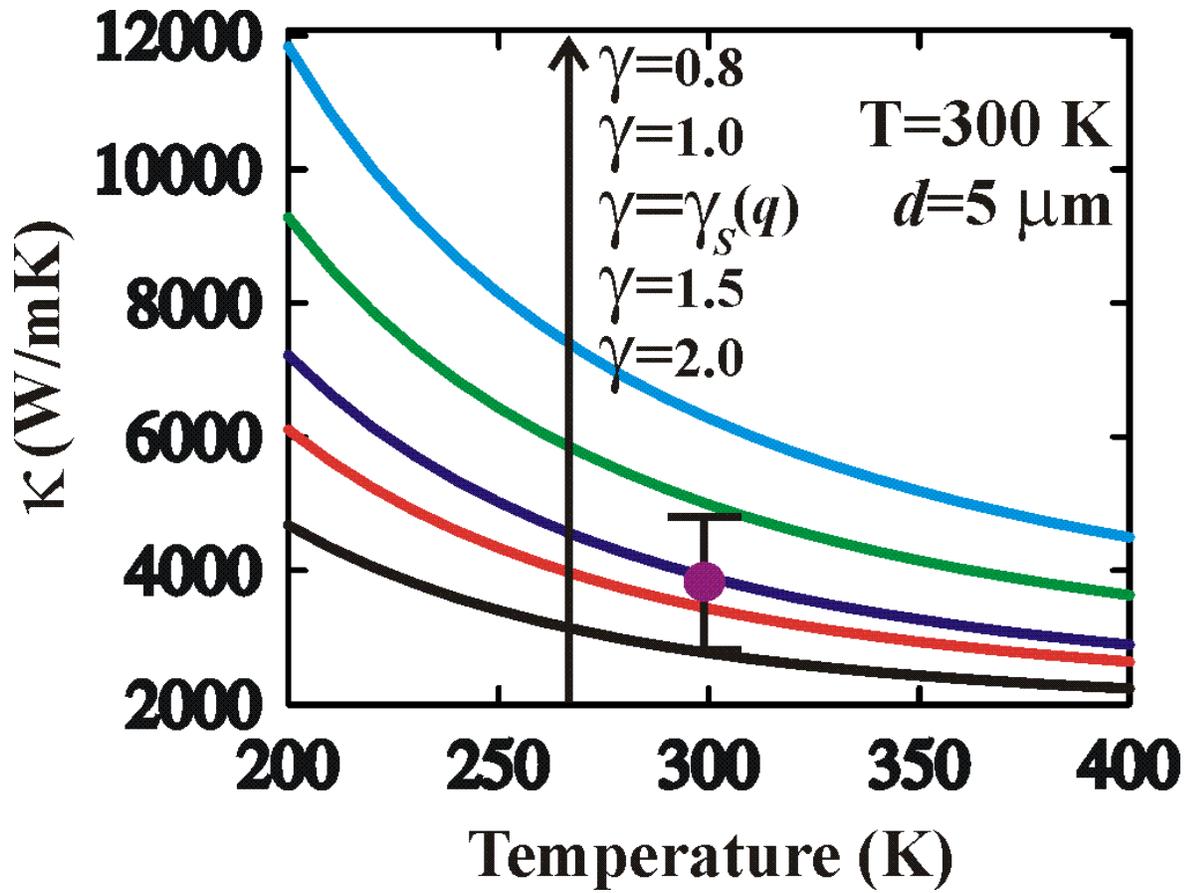

Fig. 7 of 9. Nika et.al.

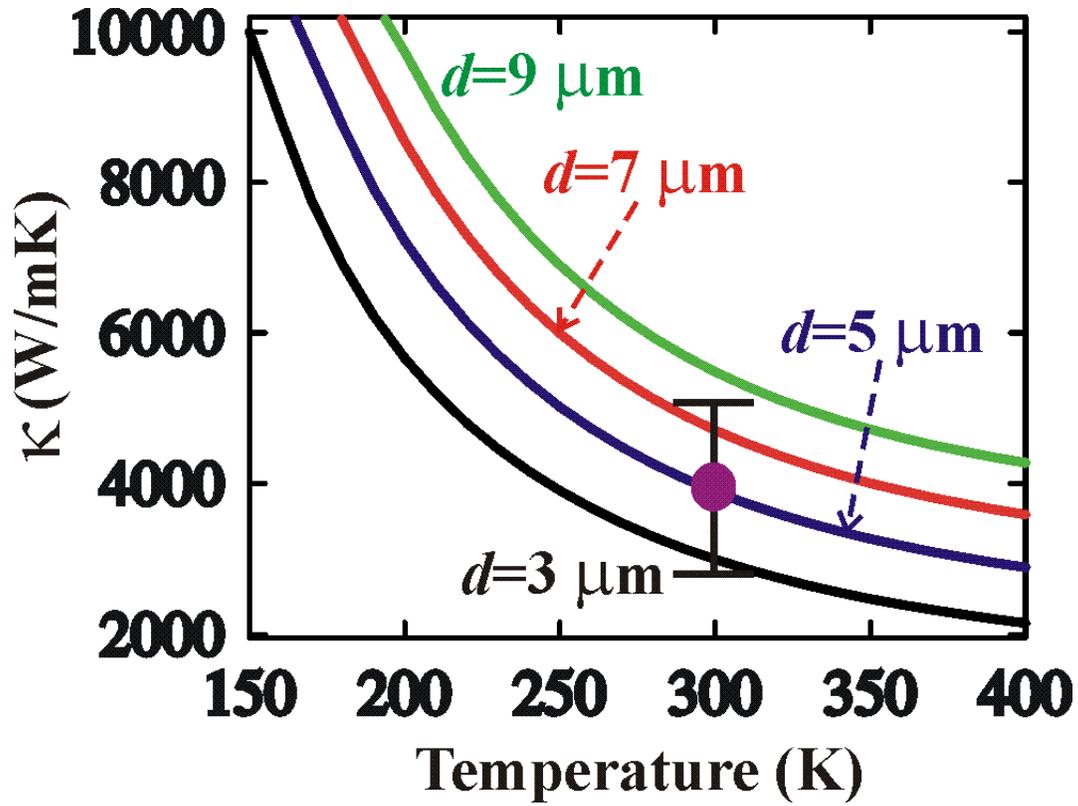

Fig. 8 of 9. Nika et.al.

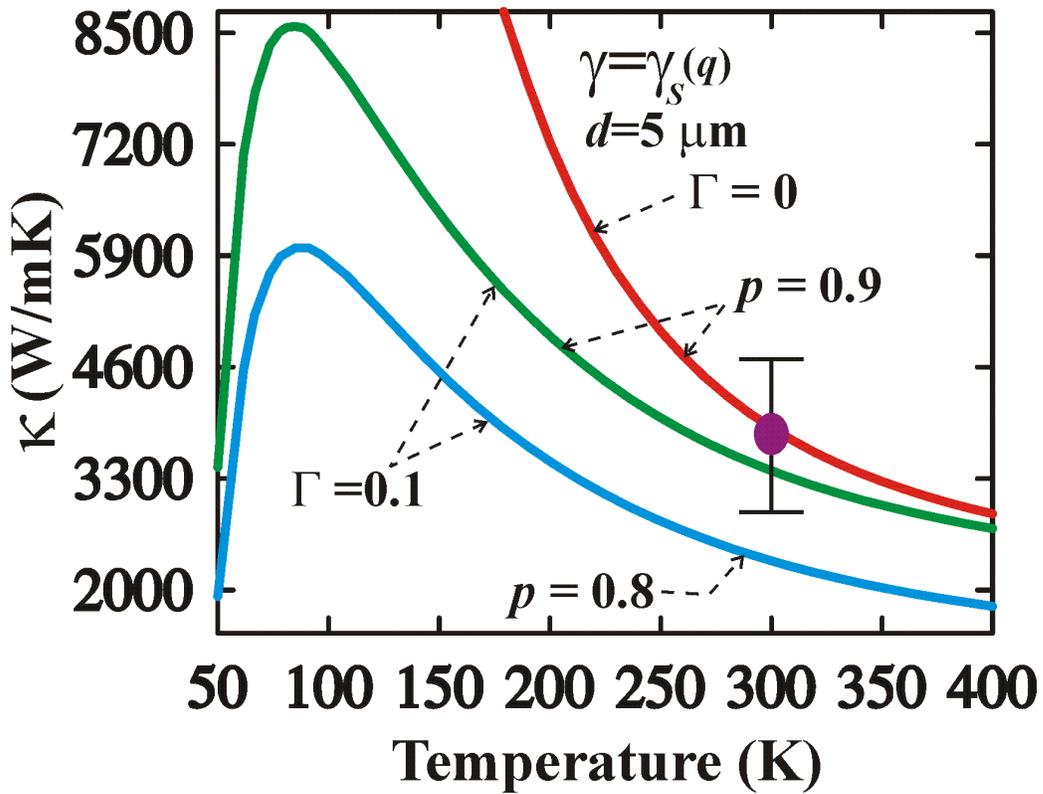

Fig. 9 of 9. Nika et.al.